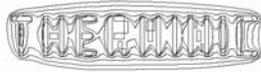

# ENABLING ELECTRONIC PROGNOSTICS USING THERMAL DATA

*Nikhil Vichare and Michael Pecht*

CALCE Electronic Products and Systems Center, University of Maryland, College Park, MD 20742, USA

## ABSTRACT

Prognostics is a process of assessing the extent of deviation or degradation of a product from its expected normal operating condition, and then, based on continuous monitoring, predicting the future reliability of the product. By being able to determine when a product will fail, procedures can be developed to provide advanced warning of failures, optimize maintenance, reduce life cycle costs, and improve the design, qualification and logistical support of fielded and future systems. In the case of electronics, the reliability is often influenced by thermal loads, in the form of steady-state temperatures, power cycles, temperature gradients, ramp rates, and dwell times. If one can continuously monitor the thermal loads, in-situ, this data can be used in conjunction with precursor reasoning algorithms and stress-and-damage models to enable prognostics. This paper discusses approaches to enable electronic prognostics and provides a case study of prognostics using thermal data.

## 1. INTRODUCTION

A significant proportion of industry world-wide is devoted to products and systems intended for operation over a long life cycle. Examples of such systems include aerospace, automotive, telecom infrastructure, oil exploration and military applications. In these systems, long-term reliability, often 10 – 20 years, is essential and maintenance opportunities are frequently limited by accessibility, as well as logistical, operational, and economic constraints. If one can assess the extent of deviation or degradation from an expected normal operating condition (i.e., the system's health), of these systems, this data can be used to meet the following powerful objectives, which include (1) providing an early warning of failure; (2) forecasting maintenance as needed to avoid scheduled maintenance and extend maintenance cycles; (3) assessing the potential for life extensions; (4) reducing the amount of redundancy; (5) providing guidance for system re-configuration and self-healing; (6) providing efficient fault detection and identification, including evidence of "failed" equipment found to function properly when re-tested (no-fault found); and (7) improving future designs and qualification methods [1] [2].

Most products and systems contain some electronics to provide functionality and performance. These electronics are often the first item of the product or system to fail [1]. While the application of health monitoring, is well established for assessment of mechanical systems, this is not the case for electronic systems. The term "diagnostics" pertains to the detection and isolation of faults or failures. "Prognostics" is the process of predicting a future state (of reliability) based on current and historic conditions. Prognostics and health management (PHM) is a method that permits the reliability of a system to be evaluated in its actual life-cycle conditions, to determine the advent of failure, and mitigate the system risks.

PHM has emerged as one of the key enablers for achieving efficient system-level maintenance and lowering life-cycle costs. In November 2002 [3], the U.S. Deputy Under Secretary of Defense for Logistics and Materiel Readiness released a policy called condition-based maintenance plus (CBM+ ) [3]. CBM+ represents an effort to shift unscheduled corrective equipment maintenance of new and legacy systems to preventive and predictive approaches that schedule maintenance based upon the evidence of need.

The importance of PHM implementation was explicitly stated in the DoD 5000.2 policy document on defense acquisition [5], which states that "program managers shall optimize operational readiness through affordable, integrated, embedded diagnostics and prognostics, and embedded training and testing, serialized item management, automatic identification technology (AIT), and iterative technology refreshment" [4]. A 2005 survey of eleven CBM programs highlighted "electronics prognostics" as one of the most needed maintenance-related features or applications, without regard for cost [5], a view also shared by the avionics industry [6].

In this paper the main approaches for prognostic implementation are categorized as (1) use of expendable devices, such as 'canaries' and fuses that fail earlier than the host product to provide advance warning of failure; (2) monitoring and reasoning of parameters that are precursors to impending failure, such as shifts in performance parameters; and (3) modeling of stress and damage in electronic parts and structures utilizing exposure conditions (e.g., usage, temperature, vibration, radiation) to compute accumulated damage. For electronic systems, in-situ data can be monitored from different sites including device, components, interconnects, boards, and systems using built-in sensors and data-buses (JTAG, I2C, etc) or external sensors. The monitored data can be used with the suitable prognostic approaches to provide the health assessment and remaining life (Figure 1). A PHM case-study using thermal data is provided.

## 2. FUSES AND CANARIES

Expendable devices such as fuses and canaries have been a traditional method of protection for structures and electrical power systems. For example, fuses and circuit breakers are used in electronic products to sense excessive current drain and to disconnect power from a product or system. Fuses within circuits safeguard against voltage transients or excessive power dissipation, and protect power supplies from shorted parts.

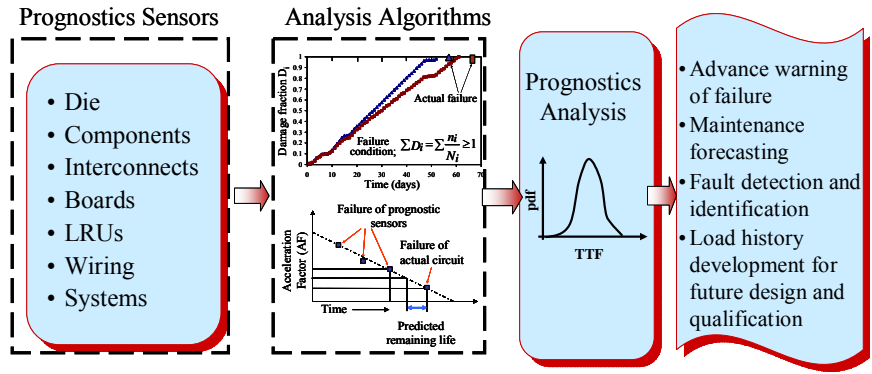

Figure 1. CALCE Framework for Electronic Prognostics

The word "canary" is derived from one of coal mining's earliest systems for warning of the presence of hazardous gas using the canary bird. Because the canary is more sensitive to hazardous gases than humans, the death or sickening of the canary was an indication to the miners to get out of the shaft. The canary thus provided an effective early warning of catastrophic failure by providing advance warning that was easy to interpret.

Figure 2 shows the failure distribution of the actual product and the canary devices. Under the same environmental and operational loading conditions, the canary health monitors wearout faster to indicate the impending failure of the actual product. Canaries can be calibrated to provide sufficient advance warning of failure (prognostic distance) to enable appropriate maintenance and replacement activities. This point can be adjusted to some other early indication level. Multiple trigger points can also be provided, using multiple canaries evenly spaced over the bathtub curve. The installation of canaries mounted on the same chip as that of the actual circuitry enables multiple sensing with different acceleration factors. Canaries with higher acceleration factor will fail faster than the others. Figure 3 shows the failure points of canaries on a plot of acceleration factor vs. time. The failure points represent failure of different canaries that are pre-designed with different acceleration factors to get different failure time. Acceleration factor "one" on the line corresponds to the failure of the actual circuit. Given the plot, the remaining life of the actual circuit can be predicted, which is the time difference between the points with an acceleration factor equal to "one" and a time where the prediction is needed.

Fuses and canary devices can be used to provide advance warning of failure due to specific wearout failure mechanisms. Mishra, et al., [7] studied the applicability of semiconductor-level health monitors by using pre-calibrated cells (circuits) located on the same chip with the actual circuitry. The prognostics cell approach has been commercialized by Ridgetop Group (known as Sentinel Semiconductor$^{TM}$ technology) to provide an early-warning sentinel for upcoming device failures [8]. The prognostic cells are available for 0.35, 0.25, and 0.18 micron CMOS processes; the power consumption is approximately 600 microwatts. The cell size is typically 800 $\mu m^2$ at the 0.25 micron process size. Currently, prognostic cells are available for semiconductor failure mechanisms such as electrostatic discharge (ESD), hot carrier, metal migration, dielectric breakdown, and radiation effects.

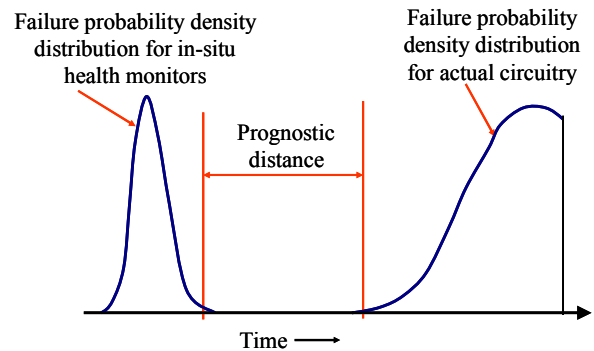

Figure 2. Advance warning of failure using canary structures

The time to failure of these prognostic cells can be pre-calibrated with respect to the time to failure of the actual product. Because of their location, these cells contain and experience substantially similar dependencies as does the actual product. These stresses that contribute to degradation of the circuit include voltage, current, temperature, humidity, and radiation. Since the operational stresses are the same, the damage rate is expected to be the same for both the circuits. However, the prognostic cell is designed to fail faster through increased stress on the cell structure by means of scaling.

Scaling in fuses and canaries can be achieved by controlled increase of the current density inside the cells. With the same amount of current passing through both circuits, if the cross-sectional area of the current-carrying paths in the cells is decreased, a higher current density is achieved. Further control in current density can be achieved by increasing the voltage level applied to the cells. A combination of both of these techniques can also be used. Higher current density leads to higher internal (joule) heating, causing greater stress on the cells. When a current of higher density passes through the cells, they are expected to fail faster than the actual circuit [7].

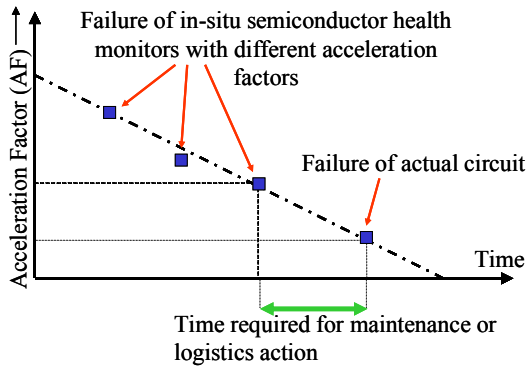

Figure 3. Multiple canaries calibrated to different levels of degradation

## 3. MONITORING PRECURSORS TO FAILURE

A failure precursor is an event that signifies impending failure. A precursor indication is usually a change in a measurable variable that can be associated with subsequent failure. For example, a shift in the output voltage of a power supply might suggest impending failure due to damaged feedback regulator and opto-isolator circuitry. Failures can then be predicted by using a causal relationship between a measured variable that can be correlated with subsequent failure.

A first step in monitoring precursors to failure is to select the life-cycle parameters to be monitored. Parameters can be identified based on factors that are crucial for safety, that are likely to cause catastrophic failures, that are essential for mission completeness, or that can result in long downtimes. Selection can also be based on knowledge of the critical parameters established by past experience and field failure data on similar products and on qualification testing. More systematic methods, such as failure mode mechanisms and effects analysis (FMMEA) [9], can be used to determine parameters that need to be monitored.

It is also necessary to develop a reasoning algorithm to correlate the change in the precursor variable with the impending failure. This characterization is typically performed by measuring the precursor variable under an expected or accelerated usage profile. Based on the characterization, a model is developed -- typically a parametric curve-fit, neural-network, Bayesian network, or a time-series trending of a precursor signal(s).

For a fielded product with highly varying usage profiles, an unexpected change in the usage profile could result in a different (non-characterized) change in the precursor signal. If the precursor reasoning model is not characterized to factor in the uncertainty in life-cycle usage and environmental profiles, it may provide false alarms. Additionally, it may not always be possible to characterize the precursor signals under all possible usage scenarios (assuming they are known and can be simulated). This approach can be improved by using precursors that relate to the failure mechanisms, instead of precursors purely based on observation.

The dependence of training to build the precursor reasoning model can be explained using a simple example regarding use of a computer for office work and engineering analysis. A precursor analysis tool may have been trained while it was used for office applications such as word processing and data entry that did not involve any data processing. The precursor is chosen to be the case temperature of the microprocessor. This training will result in identification of a temperature signal with a narrow band with a relatively low mean value. This situation is analogous to the creation of a control chart during manufacturing. When the same computer is now used for engineering analysis that utilizes the full capability of the processor, the average temperature of the processor case and the variations in the temperature is likely to increase; these conditions will be seen as an aberration from normal usage resulting in false alarm. On the other hand, if the training condition is reversed, then even an unnatural temperature rise during word processing application that is caused by real degradations (e.g., processor die attach delamination, fan capacity loss, heat sink fouling) will not be detected as a cause for concern from the temperature signal.

This situation leads us to a few basic tenets of the physics of failure based design principles. The life cycle environmental condition need to be established and continually updated to take advantage of the prognostics built into a system. Also, we need to first know the failure mechanisms that are likely to cause the degradations that can lead to eventual failures in the systems. It is necessary to understand that there are different failure signatures for different failure mechanisms and without knowledge of the failure mechanisms to monitor; it is impractical to expect success in implementation of PHM.

## 4. MONITORING ENVIRONMENTAL AND USAGE LOADS

The life-cycle environment of a product consists of manufacturing, storage, handling, operating and non-operating conditions. The life-cycle loads, either individually or in various combinations, may lead to performance or physical degradation of the product and reduce its service life [2]. The extent and rate of product degradation depends upon the magnitude and duration of exposure (usage rate, frequency, and severity) to such loads. If one can measure these loads in-situ, the load profiles can be used in conjunction with damage models to assess the degradation due to cumulative load exposures.

The assessment of the impact of life-cycle usage and environmental loads on electronic structures and components was studied by Ramakrishnan and Pecht [10]. This study introduced the life consumption monitoring (LCM) methodology, which combines in-situ measured loads with physics-based stress and damage models for assessing the life consumed.

The application of the LCM methodology to electronics PHM was illustrated with two case studies [10]. Temperature and vibrations were measured in-situ on the board in the application environment. Using the monitored environmental data, stress and damage models were developed and used to estimate consumed life. The LCM methodology predicted remaining life with acceptable accuracy. The application of this approach to legacy systems was demonstrated by Mathew, et al., [11] and Shetty, et al. [12].

Vichare et al [2] [13] [14] proposed methods for monitoring and recording the loads in-situ. Methods were proposed for embedding the data reduction and load parameter extraction algorithms in to the sensor modules to enable reduction in on-board storage space, low power consumption, and uninterrupted data collection over longer durations. As shown in Figure 4, a time-temperature signal is monitored in-situ using sensors, and further processed to extract cyclic temperature range ($\Delta T$), cyclic mean temperature ($T_{mean}$), ramp rate ($dT/dt$), and dwell time ($t_D$) using embedded load extraction algorithms. The extracted load parameters are stored in appropriate bins to achieve further data reduction. The binned data is downloaded to estimate the distributions of the load parameters for use in damage assessment, remaining life estimation, and the accumulation of the products use history.

The methodology presented above for load monitoring and analysis is applied to the notebook computer, which was previously thermally characterized by Vichare et al. 2004 [14]. While a comprehensive health monitoring plan may involve multiple life cycle conditions, such as humidity, vibration, shock, radiation, and contamination, the present study focuses on temperature. A list of variables (not exhaustive) for monitoring the entire laptop is proposed in Table 1. Internal temperatures on CPU heat sink temperature, hard disk drive (HDD) temperature, ambient temperature (in vicinity of laptop), were dynamically monitored along with percentage of CPU utilized and fan condition (on/off), in-situ during all phases of the life cycle, including usage, storage, and transportation.

Figure 5a shows the raw sensor data, namely absolute temperature-time history for the heat sink and HDD, the corresponding temperature cycles amplitudes, and CPU usage. The raw sensor absolute temperature data shown in Figure 5a was converted into a sequence of peaks and valleys using the ordered overall range method. The sequence of peaks and valleys were processed using a 3-parameter Rainflow method to identify the complete and half cycles and extract the amplitude, mean temperature, and ramp rate of each cycle. The resulting distribution of cyclic temperature ranges and mean temperatures on the heat sink and HDD are shown in Figure 5b and Figure 5c respectively. These data can be used with stress and damage models to asses the degradation of product and estimate the remaining life of the product. Along with the temperatures the usage of the product was recorded by logging the percentage CPU usage and the operation of the CPU cooling fan. An example of CPU usage monitoring is shown in Figure 5d with CPU usage and corresponding heat sink temperature.

Table 1. Environmental, usage and performance parameters for PHM of notebook computers

| Parameters | Examples |
| --- | --- |
| Environmental and Usage | - Temperatures of microprocessor, hard disk drive, video card, RAM etc.<br>- Vibrations/shock in application and handling, disk spin<br>- Processor usage, memory usage, processor queue length, Cache fast reads/second etc.<br>- Power cycles, number of on/off<br>- Hard disk and monitor on/off<br>- Strain in mother board flexing and torsion during handling and due to ageing<br>- Pressure on keyboard and buttons<br>- Force on latch, hinge, and connectors<br>- Humidity and radiation exposure |
| Performance and system setup | - Fan ON/OFF, Fan speed<br>- CPU core voltage, CPU I/O voltage<br>- Hard disk parameters: spin-up time, flying head height, ECC count, data transfer rate<br>- Power management settings (power schemes, hibernate, and stand-by settings) |

In Figure 5b, the CPU heat sink temperature was found to be 13$^o$C and 8$^o$C lower than its maximum rating over 90% and 95% of the monitored time period, respectively. This highlights the potential conservativeness of thermal management solutions optimized based on worst-case operating conditions that rarely occur. Such findings could contribute to the design of more sustainable, least-energy consumption thermal management solutions.

Approximately 97% of the temperature cycles experienced by either the CPU heat sink or hard disk drive had amplitude of less than 5$^o$C. However, the maximum temperature cycle amplitudes measured were found to exceed those specified by environmental standards for computer and consumer equipment. Such a discrepancy between standardized and actual conditions provides a strong motivation for monitoring actual product application environments.

The effects of power cycles, usage history, CPU computing resources usage, and external thermal environment on peak transient thermal loads were characterized. It was found that the product thermal operating envelope differed considerably from the operating conditions specified by environmental standard IPC SM-785 [15] and from the worst-case operating conditions

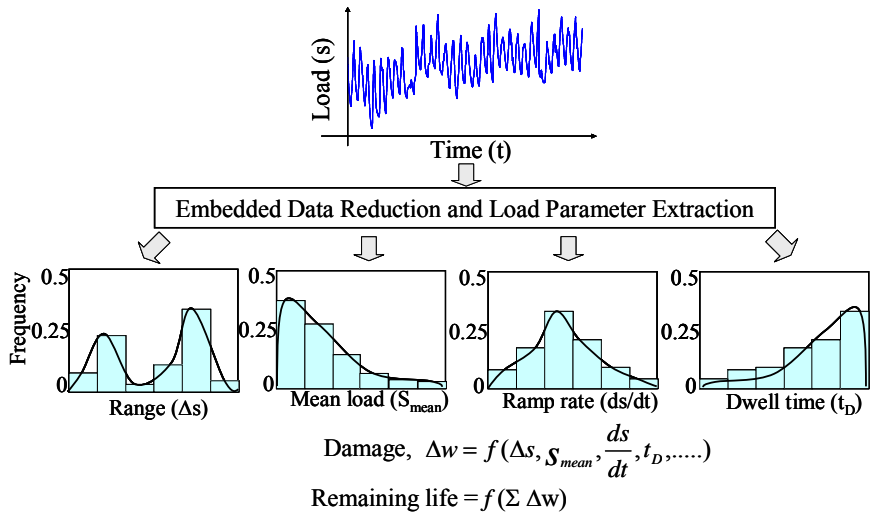

Figure 4. Load parameter extraction for health assessment and prognostics

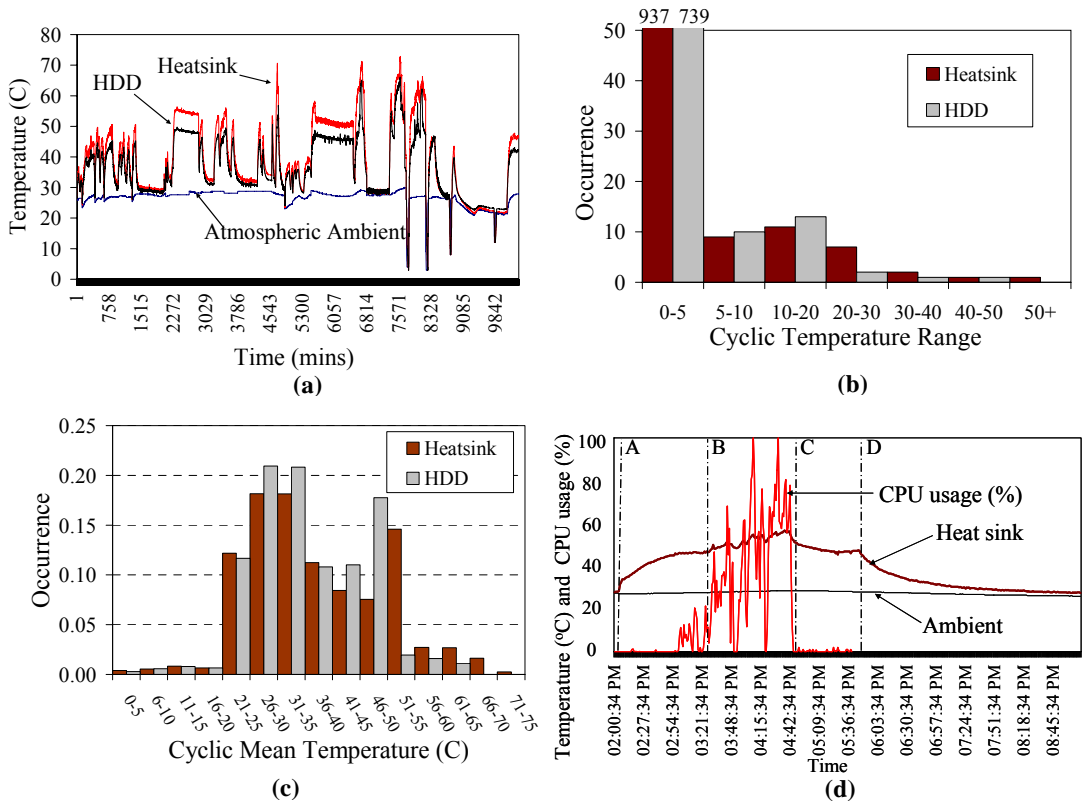

Figure 5. Monitoring and analysis of environmental and usage conditions in a notebook computer: (a) Example illustration of raw sensor data, (b) Distribution of absolute temperatures on heatsink and HDD, (c) Distribution of temperature cycles on heatsink and HDD

assumed by the vendor, in terms of absolute temperatures, temperature cycle amplitudes, and temperature ramp rates. The monitored life cycle temperature data could also be used in physics based stress and damage models, to provide both damage estimation and remaining life prediction due to specific failure mechanisms influenced by temperature. The measured data could also be used to determine the stress levels to be imposed in accelerated testing, refining product specifications, and setting product warranties.

## 5. PHM INTEGRATION

Implementing an effective PHM strategy for a complete product or system may require integrating different prognostic health monitoring approaches. The first step is an analysis to determine the weak link(s) in the system based on the potential failure modes and mechanisms to enable a more focused monitoring process. Once the potential failure modes, mechanisms, and effects (FMMEA) have been identified, a combination of canaries, precursor reasoning, and life-cycle damage modeling may be necessary, depending on the failure attributes. In fact, different approaches can be implemented based on the same sensor data.

For example, operational loads, such as temperature, voltage, current, and acceleration can be used with damage models to calculate the susceptibility to electromigration between metallizations, and thermal-fatigue of interconnect, plated-thru holes, die-attach etc. Also, the processor usage, current, and CPU temperature data can be used to build a statistical model that is based on the correlations between these parameters. This model can be appropriately trained to detect thermal anomalies and identify sign for certain transistor degradation.

Future electronic system designs will integrate sensing and processing modules that will enable in-situ PHM. Advances in sensors, microprocessors, compact non-volatile memory, battery technologies, and wireless telemetry have already enabled the implementation of sensor modules and autonomous data loggers. For in-situ health monitoring, integrated, miniaturized, low-power, reliable sensor systems operated using portable power supplies (such as batteries) are being developed. These sensor systems have self-contained architecture requiring minimum or no intrusion into the host product in addition to specialized sensors for monitoring localized parameters. Sensors with embedded algorithms will enable fault detection, diagnostics, and remaining life prognostics that would ultimately drive the supply chain. The prognostic information will be linked via wireless communications to relay needs to maintenance officers and automatic identification techniques (RFID being the most common current example) will be used to locate parts in the supply chain--all integrated through a secure web portal to acquire and deliver replacement parts quickly on an as-needed basis.